\documentclass[sigplan,10pt]{acmart}

\usepackage{tikz}
\usepackage{amsmath}
\usepackage{xspace}
\usepackage{xcolor}
\usepackage{enumerate}
\usepackage{siunitx}
\usepackage[shortlabels]{enumitem}
\usepackage{todonotes}

\newcommand*{\eg}{e.g.,\@\xspace}

\newcommand*{\vs}{vs.\@\xspace}
\newcommand*{\cf}{cf.\@\xspace}

\newcommand{\sasos}{single-address-space operating system\xspace}

\newcommand{\Sasoses}{Single-address-space operating systems\xspace}

\newcommand{\SASOS}{Single-Address-Space OS\xspace}
\newcommand{\SASOSes}{Single-Address-Space OSes\xspace}

\newcommand{\SASOSesfull}{Single-Address-Space Operating Systems\xspace}

\newcommand{\cow}{copy-on-write\xspace}

\newcommand{\CoW}{Copy-on-Write\xspace}

\newcommand{\CoA}{Copy-on-Access\xspace}

\newcommand{\CoPA}{Copy-on-Pointer-Access\xspace}

\newcommand{\vfork}{\texttt{vfork}\xspace}
\newcommand{\fork}{\texttt{fork}\xspace}
\newcommand{\sfork}{\si{\micro Fork}\xspace}
\newcommand{\bfsfork}{\si{\textbf{\micro Fork}}\xspace}
\newcommand{\itsfork}{\si{\textit{\micro Fork}}\xspace}
\newcommand{\process}{\si{\micro process}\xspace}
\newcommand{\itprocess}{\si{\textit{\micro process}}\xspace}
\newcommand{\bfprocess}{\si{\textbf{\micro process}}\xspace}
\newcommand{\processes}{\si{\micro \text{{processes}}}\xspace}
\newcommand{\itprocesses}{\si{\textit{$\mu$processes}}\xspace}
\newcommand{\bfprocesses}{\si{\textbf{\micro processes}}\xspace}

\DeclareRobustCommand{\BC}[1]{\tikz[baseline=(char.base)]{
	\node[shape=circle,draw,inner sep=0.15pt, fill={rgb:red,197;green,0;blue,11}] (char) {\textcolor{white}{#1}};}}

\begin{document}
\copyrightyear{2025}
\acmYear{2025}
\setcopyright{cc}
\setcctype{by}
\acmConference[SOSP '25]{ACM SIGOPS 31st Symposium on Operating Systems
Principles}{October 13--16, 2025}{Seoul, Republic of Korea}
\acmBooktitle{ACM SIGOPS 31st Symposium on Operating Systems Principles
(SOSP '25), October 13--16, 2025, Seoul, Republic of Korea}
\acmDOI{10.1145/3731569.3764809}
\acmISBN{979-8-4007-1870-0/2025/10}
\date{}


\title{\bfsfork: Supporting POSIX \fork Within a \SASOS}

\author{\rm John Alistair Kressel}
\affiliation{%
  \institution{The University of Manchester}
\city{Manchester}
  \country{United Kingdom}}
\author{\rm Hugo Lefeuvre}
\affiliation{
\institution{The University of British Columbia}
\city{Vancouver, BC}
\country{Canada}}
\author{\rm Pierre Olivier}
\affiliation{
\institution{The University of Manchester}
\city{Manchester}
\country{United Kingdom}}

\begin{abstract}

\Sasoses have well-known lightweightness benefits that result from their central design idea: the kernel and applications share a unique address space.
This model makes these operating systems (OSes) incompatible by design with a large class of software: multiprocess POSIX applications.
Indeed, the semantics of the primitive used to create POSIX processes, \fork, are inextricably tied to the existence of multiple address spaces.

Prior approaches addressing this issue trade off light\-weightness, compatibility and/or isolation.
We propose \itsfork, a \sasos design supporting POSIX \fork on modern hardware without compromising on any of these key objectives.
\sfork emulates POSIX processes (\processes) and achieves \fork by creating for the child a copy of the parent \process' memory at a different location within a single address space.
This approach presents two challenges: relocating the child's absolute memory references (pointers), as well as providing user/kernel and \processes isolation without impacting lightweightness.
We address them using CHERI.
We implement \sfork and evaluate it upon three real-world use-cases: Redis snapshots, Nginx multi-worker deployments, and Zygote FaaS worker warm-up.
\sfork outperforms previous work and traditional monolithic OSes on key lightweightness metrics by an order of magnitude, e.g. it can offer a \fork-bound FaaS function throughput 24\% higher than that of a monolithic OS, and can \fork a \process in 54 \si{\micro s}, 3.7$\times$ faster than a traditional \fork.
\end{abstract}

\maketitle 
\pagestyle{empty} 

\section{Introduction}

A \sasos (SASOS) colocates the OS kernel together with all user programs within a single address space to achieve \emph{lightweightness} benefits: performance improvements through optimized context switches and swift security domain transitions, as well as low memory footprint, among others.
Early SASOSes such as Opal~\cite{koldinger_architecture_1992} and Mungi~\cite{heiser_mungi_1998} leveraged this design to ease communication and coordination across processes.
More recently, the rise of virtualization has enabled customizable SASOSes which simplify or completely eliminate barriers between user and kernel code~\cite{kuenzer_unikraft_2021, madhavapeddy_unikernels_2013}, to achieve large performance and resource usage improvements.
Due to their lightweight characteristics, SASOSes are growing in popularity in many areas where high performance and low resource consumption is required, including Function as a Service (FaaS)~\cite{cadden_seuss_2020, fingler_usetl_2019, noauthor_kraftcloud_nodate}, edge computing~\cite{cozzolino_fades_2017, wang_energy-efficient_2019, morabito_consolidate_2018} and confidential computing~\cite{GramineTDX}.

Still, despite this growing popularity, SASOSes struggle to find mainstream adoption.
\emph{A key obstacle to their widespread deployment is their lack of support for a wide range of software: multiprocess POSIX applications~\footnote{We use the POSIX definition of process, which implies a separate address space. However, we note that the term "process" long predates POSIX.}}~\cite{zhang_kylinx_2018, noauthor_porting_nodate, noauthor_unikernels_nodate, kuo_linux_2020, tsai_cooperation_2014}.
These applications rely on the primary process creation mechanism in POSIX systems, \fork.
\fork is an old and ubiquitous mechanism that creates a new (child) process by copying the memory and system resources of the calling process (parent).
Many popular applications such as OpenSSH~\cite{provos_preventing_2003}, Redis~\cite{noauthor_redis_2024}, Nginx~\cite{noauthor_nginx_2024}, Apache httpd~\cite{noauthor_welcome_nodate} and Qmail~\cite{bernstein_qmail_2007} are designed to take advantage of \fork for purposes such as concurrency, isolation, and on-demand resource duplication.
Unfortunately, the semantics of \fork make it fundamentally hard to support in a SASOS: with \fork, each child process is created in a new address space, whereas a true SASOS can only have, \textit{by definition}, a single one.

An ideal approach to support \fork within SASOSes should accomplish all the following objectives.
First, \fork support must not compromise the lightweightness of SASOSes.
In particular, it should not re-introduce multiple address spaces, since a single address space is critical to the performance advantages which have been exploited for fast IPC~\cite{directIPC, noauthor_ctsrd-chericomsg_2024} and I/O~\cite{ix_dataplane}, among others.
Second, the inter-process \emph{isolation} properties obtained with traditional \fork, confining each process within its own address space, must be preserved in a single-address-space implementation.
SASOS \fork support must also be \emph{transparent} for applications, i.e. it must exhibit the same semantics as traditional \fork.
Finally, SASOS \fork must also \emph{perform} at least as well as traditional \fork, exhibiting the same memory usage and performance (e.g. \cow) characteristics as typical POSIX OSes.

Over the past decades, several solutions to this problem have been proposed, yet all trade off on at least one of the aforementioned key objectives.
In the 1990s, early SASOSes used segment-relative addressing to \fork processes~\cite{wilkinson_compiling_1993, heiser_mungi_1998}, avoiding the need for relocations by making all memory accesses relative to a base register.
However, segment-relative addressing is poorly supported in modern toolchains and ISAs, and bringing it back would imply significant modifications to a large amount of systems software including compilers, JIT runtimes, or handwritten assembly code.
Other efforts abandon isolation~\cite{OSV_EXECVE}, which raises security concerns.
Yet others abandon transparency by manually porting applications to avoid calling \fork~\cite{OSV_NGINX_PATCH}, something that cannot be done with most \fork use cases without a significant per-application engineering effort.
Other attempts have leveraged virtualization hardware to duplicate the entire OS and application~\cite{lupu_nephele_2023, zhang_kylinx_2018, tsai_cooperation_2014}, which sacrifices lightweightness by introducing multiple address spaces.

We present the design and implementation of \sfork to address the need for efficient \fork support within a SASOS.
Unlike existing approaches, \sfork is a true single-address-space solution which does not compromise on lightweightness, isolation, transparency, or performance.
In essence, \sfork emulates POSIX processes (\processes) and supports POSIX \fork by copying the memory used by a parent \process to a different location within a single address space, for use by the child \process.
This approach leads to two challenges: 1) identifying in the child absolute memory references pointing to the parent's area, and relocating them to the child's; and 2) enforcing isolation between \processes, as well as between \processes and the kernel, without compromising on lightweightness.
To solve these challenges, we leverage two modern technologies: hardware memory tagging to identify and relocate absolute memory references to the correct location within the child's area, and intra-address-space memory protection to enforce isolation across \processes and the OS kernel.
Building on this, \sfork proposes novel optimized \cow strategies.

We implement a prototype of \sfork using CHERI~\cite{woodruff_cheri_2014, watson_cheri_2015} on the ARM Morello platform~\cite{arm_ltd_arm_2022} using as a basis the Unikraft SASOS~\cite{kuenzer_unikraft_2021}.
CHERI's tagged memory allows us to relocate absolute memory references, and its tightly bounded hardware memory capabilities let us enforce isolation across \processes and the OS kernel.
We evaluate our \sfork prototype on microbenchmarks and real-world, unmodified, \fork-based applications: Redis snapshots, Nginx multi-worker deployments, and a Micro\-Python-based serverless computing framework warming up Zygote~\cite{oakes_sock_2018} workers.
This lets us showcase key metrics such as \fork latency, memory usage, and overall application performance.
We compare \sfork to a classical POSIX \fork on a CHERI-enabled FreeBSD, and a unikernel \fork implementation leveraging virtual machines (VMs), Nephele~\cite{lupu_nephele_2023}.
We show that \sfork outperforms these approaches by an order of magnitude on key metrics: \sfork can \fork a \process in 54 \si{\micro s}, 3.7$\times$ faster than a traditional \fork on FreeBSD, and 198$\times$ faster than Nephele's VM-based approach, while reducing the memory usage by respectively 2.2$\times$ and 12.3$\times$.

Overall, this paper contributes:
\begin{itemize}[leftmargin=0.5cm,noitemsep]
\item An analysis and breakdown of the problem of supporting \fork in SASOSes, along with a systematic perspective on prior attempts (\S\ref{sec:background} and \S\ref{sec:design}).
\item The design of \sfork, a novel approach to implement \fork in SASOSes,	which relies on recent advances in hardware-based isolation technologies (\S\ref{sec:design}).
\item The open-source implementation (\S\ref{sec:implementation}) and evaluation (\S\ref{sec:evaluation}) of \sfork with CHERI and Unikraft.
\end{itemize}

\section{Background} \label{sec:background}
In this section we examine the use of \fork in modern software.
Based on this, we discuss the lack of \fork support in SASOSes, and its impact.
We then contextualize existing attempts to support \fork in SASOSes, and position our proposed contribution, \sfork.
Finally, we provide background on CHERI, which we use to implement \sfork.

\subsection{\fork: A Pervasive Software Design Primitive} \label{forktaxonomy}
\fork is the oldest process creation mechanism in POSIX operating systems, tracing its roots back to Project Genie in the 1960s~\cite{nyman_notes_2016}.
When a process calls \fork{}, the OS kernel creates a copy of the process' memory, state, and system resources such as sockets and open file descriptors into a new isolated address space.
The newly created process is referred to as the \emph{child}, and the \fork caller as the \emph{parent}.
Although conceptually simple, real-world POSIX \fork comes with optimizations such as \cow, and many features which allow customization of its copying and protection behavior~\cite{baumann_fork_2019}.
Most importantly, the semantics of \fork imply that it is responsible for both creating new processes \emph{and} creating new address spaces.
This makes \fork a mechanism that imposes strong constraints on OS design~\cite{baumann_fork_2019}, and makes it fundamentally difficult to implement in SASOSes which have, by definition, only one address space.

Still, \fork remains a very popular software design primitive, whose uses typically match the following patterns:

\begin{enumerate}[label={\textbf{\textit{(U\arabic*)}}}, leftmargin=0.8cm,noitemsep]
\item \label{U1} \textbf{\fork + \texttt{exec} to start a new program.} Examples include running an executable via Bash.

\item \label{U2} \textbf{\fork for concurrency.} Web servers such as Nginx~\cite{noauthor_nginx_2024} and Apache~\cite{noauthor_welcome_nodate} use \fork to create additional worker processes that handle requests concurrently.

\item \label{U3} \textbf{\fork for privilege separation.} Privilege-separated software such as OpenSSH~\cite{provos_preventing_2003} and qmail~\cite{bernstein_qmail_2007} leverage \fork to isolate trusted and untrusted application parts.

\item \label{U4} \textbf{\fork to leverage \cow.} The Redis~\cite{noauthor_redis_2024} key-value store uses \fork to create an instant snapshot of an in-memory database with \cow and saves it to the disk, concurrently with the main database process that continues to handle requests.

\item \label{U5} \textbf{\fork to reduce startup times.} Testing frameworks such as fuzzers use \fork to avoid the cost of setup for each exploration~\cite{noauthor_googleafl_2024, nagy_full-speed_2019, noauthor_googlehonggfuzz_2024, noauthor_libfuzzer_nodate, xu_designing_2017} and improve efficiency.
Latency-sensitive applications such as Function as a Service (FaaS) use \fork to quickly start function instances~\cite{ao_faasnap_2022}.
 Android applications are started through \fork called by a root Zygote process~\cite{ANDROID_ZYGOTE}.

 \item \label{U6} \textbf{\fork to daemonize.} By leveraging concurrency, applications such as web servers can create detached processes which run independently in the background.
\end{enumerate}

Applications may also use a combination of all, showing how \fork is vital to run popular applications.
We found that the 46\% of the 50 most popular C repositories on GitHub and 50\% of 50 most popular Debian packages reported by popcon~\cite{popcon}, use \fork.

\subsection{\SASOSesfull} \label{subsec:background:sasoses}
\Sasoses were first proposed in the early 1990s, enabled by the spread of 64-bit processors which made it possible to colocate the kernel and all applications within a single address space~\cite{chase_sharing_1994, roscoe_linkage_1994, koldinger_architecture_1992}.
Initially, SASOSes were introduced to simplify data sharing and communication across processes, since each has the same flat view of memory.
Additionally, SASOSes removed some of the overhead of address translation through using a unique mapping of virtual to physical addresses~\cite{koldinger_architecture_1992, heiser_mungi_1998}.
Over the years, many different types of SASOSes have been explored including exokernels~\cite{exokernel} which simplify OS abstractions by allowing applications to manage resources to suit their needs, single level store OSes~\cite{levy2014capability} which enable all data to be addressed directly by applications, dataplane OSes~\cite{ix_dataplane} which simplify I/O operations, software-isolated OSes~\cite{hunt_singularity_2007, boos_theseus_2020} which leverage type and memory safety to reduce process isolation overheads, and unikernels~\cite{kuenzer_unikraft_2021, madhavapeddy_unikernels_2013, lefeuvre_flexos_2022, kressel_software_2023, sartakov_cubicleos_2021, olivier_binary-compatible_2019} which remove barriers between user and kernel and execute all code at the same privilege level to speed up system calls.

Despite this diversity of designs, all SASOSes aim to improve \textit{\textbf{lightweightness}}, i.e. the high performance and low resource consumption benefits stemming from the minimalist nature of this OS model.
These benefits include in particular fast IPCs, low-latency security domain switches (context switches across processes, user/kernel transitions), and low memory footprint. 
They are obtained through the concise and specialized designs of SASOSes.
The primary means for achieving this lightweightness is through the single address space, removing the need for page table switches and the associated costly hardware cache flushes~\cite{tlbshootdowns} which result from context switches on multi-address-space OSes.

Although SASOSes are gaining traction in domains such as cloud computing~\cite{noauthor_kraftcloud_nodate} and confidential computing~\cite{GramineTDX, MONZA, Duy2025}, a key obstacle to their widespread deployment is their inability to efficiently support multiprocess applications using POSIX \fork~\cite{chase_sharing_1994}, which limits their compatibility with many popular applications (\eg 50\% of the 50 most popular Debian packages).
We address this shortcoming with \sfork.

\subsection{Existing Approaches to \fork in SASOSes}

\begin{table}
    \footnotesize
    \centering
    \caption{Comparison of \sfork with other SASOS \fork systems. \textit{SAS} = Single Address Space; \textit{SC} = Self-Contained (no infrastructure changes required); \textit{Seg} = Segment-relative; \texttt{f}+\texttt{e} = \fork + \texttt{exec}.}
    \begin{tabular}{c|c|c|c|c|c|c}
         System&  SAS&  Isolation&  SC&  IPCs&  Seg & \texttt{f}+\texttt{e} only\\
         \hline
         Angel~\cite{wilkinson_compiling_1993}&  \textbf{Yes}&  \textbf{Yes}&  \textbf{Yes}&  \textbf{Fast}&  Yes & No\\
         Mungi~\cite{heiser_mungi_1998}&  \textbf{Yes}&  \textbf{Yes}&  \textbf{Yes}&  \textbf{Fast}&  Yes & No\\
         Nephele~\cite{lupu_nephele_2023}&  No&  \textbf{Yes}&  No&  Med&  \textbf{No} & \textbf{No}\\
         KylinX~\cite{zhang_kylinx_2018}&  No&  \textbf{Yes}&  No&  Med&  \textbf{No} & \textbf{No}\\
         Graphene~\cite{tsai_cooperation_2014} & No & \textbf{Yes} & No& Med & \textbf{No} & \textbf{No}\\
         Graphene SGX~\cite{tsai_graphene-sgx_2017} & No & \textbf{Yes} & No& Slow & \textbf{No} & \textbf{No}\\
         Iso-Unik~\cite{li_iso-unik_2020}&  No&  \textbf{Yes}&  \textbf{Yes}&  Med&  \textbf{No} & \textbf{No}\\
         OSv~\cite{kivity_osvoptimizing_2014} & \textbf{Yes} & No & \textbf{Yes} & \textbf{Fast} & \textbf{No} & Yes\\
         Junction~\cite{Junction_2024} & \textbf{Yes} & No & No & Med & \textbf{No} & Yes \\
        \textbf{This work:} \sfork & \textbf{Yes}& \textbf{Yes}& \textbf{Yes}& \textbf{Fast}& \textbf{No} & \textbf{No}\\
    \end{tabular}
    \centering
    \label{tab:compare_colutions}
\end{table}

Over the past decades, several solutions have been proposed to support \fork in SASOSes.
We categorize them in Table~\ref{tab:compare_colutions}.

\paragraph{Pioneer SASOSes and Segment-Relative Addressing.}
\label{sec:segmentation}
Early SASOSes such as Angel~\cite{wilkinson_compiling_1993} and Mungi~\cite{heiser_mungi_1998} leveraged segment-relative addressing to implement \fork to avoid the need for relocations upon \fork, all the while maintaining a single address space and thus fast IPC.
Segment-relative addressing requires that all memory references are relative to a base register.
However, modern compiler support for segment-relative addressing cannot handle the entirety of absolute memory references that would need relocation following \fork.
These references are pointers that live not only as globals in static memory but also on the stack/heap, and modern toolchains cannot generate code dereferencing a pointer loaded e.g. from the stack relatively to a register.
This means that extensive and complex compiler modifications would be required to implement such an approach on a modern system.
In addition, a large amount of systems software including handwritten assembly or JIT runtimes would further need to be adapted to work with relative addressing.
Modern hardware support for segment-relative addressing, such as with x86-64's \texttt{fs}/\texttt{gs} registers, also does not provide adequate support for all of the possible memory references without significant compiler and toolchain modifications.
Overall, the risks associated with the complexity of combining segment-relative addressing with the need for non-negligible systems software modifications make us choose alternative approaches for \sfork.

\paragraph{Modern SASOSes and \texttt{fork} + \texttt{exec} Support.}
Certain occurrences of this pattern \textbf{\textit{(U1)}} can be replaced with more modern and efficient mechanisms, such as \texttt{vfork + exec} or \texttt{spawn}.
These are easily supported in SASOSes~\cite{shen_occlum_2020, OSV_EXECVE} by loading the newly-started program (compiled with position independent code) at a free location of the address space.
This approach only works with a subset of use cases where the parent's duplicated state is not accessed by the child between \fork and \texttt{exec}.
Some of these works also do not implement any form of isolation between applications~\cite{kivity_osvoptimizing_2014, Junction_2024}, which is not acceptable from a security standpoint.

\paragraph{Modern SASOSes and the OS as a Process.}
Nephele~\cite{lupu_nephele_2023} and KylinX~\cite{zhang_kylinx_2018, zhang_kylinx_2021} both support \fork by treating the entire SASOS as a process and thus implement \fork in the hypervisor.
The result is a \fork that copies the entire OS similarly to how a standard POSIX \fork copies a process.
Similarly, Graphene~\cite{tsai_cooperation_2014} and Graphene-SGX~\cite{tsai_graphene-sgx_2017} both approach the libOS as a process, and support \fork by piggybacking it through the host's \fork.
These `OS-as-a-process' approaches come with limitations.
First, they re-introduce multiple address spaces and lose the SASOSes lightweightness benefits which come from having a single address space.
Second, this approach is not self-contained: it requires the SASOS to run under a host (such as a hypervisor, or another OS) which itself must implement a form of \fork.

\paragraph{Modern SASOSes and Page-Tables.}
Iso-Unik~\cite{li_iso-unik_2020} takes a different approach, implementing \fork within the SASOS.
However, it does so by retrofitting multiple address spaces back into a SASOS, which similarly hurts lightweightness.

\paragraph{Summary.}
None of these solutions both preserves the goal of SASOS lightweightness whilst ensuring isolation.
Although segment-relative addressing was used in the past, the reality is that in modern systems, complex modifications must be made to compilers and toolchains as well as engineering work to port handwritten assembly and JIT compilers.
Modern solutions lose the benefits of being a true single address space which is vital for benefits including fast IPC~\cite{directIPC, noauthor_ctsrd-chericomsg_2024} and I/O~\cite{ix_dataplane}. 
This motivates for a new approach that addresses these shortcomings.

\subsection{CHERI} \label{subsec:background:cheri}

Capability Hardware Enhanced RISC Instructions~\cite{watson_cheri_2015, woodruff_cheri_2014} (CHERI) extends RISC ISAs with safety features including, among others, memory tagging and intra-address-space isolation.
That isolation is achieved through hardware capabilities where each pointer used by a program is extended from 64-bits to 128-bits, with the additional bits used to embed the bounds and access permissions of the data structure it points to.
These bounds and permissions are checked and enforced at runtime by the hardware when the pointer is dereferenced and the bounds and permissions can only be \emph{decreased} and can never be increased.
Each pointer present in memory is marked as such with an unforgeable, single-bit validity tag stored separately in dedicated memory, which is cleared automatically upon any illegitimate attempt to modify the pointer or its permissions: a pointer dereference can only succeed at runtime if the corresponding tag is valid.
\emph{Beyond its security applications, the tag also allows easy pointer tracking in memory at runtime~\cite{CORNUCOPIA, CHERIPICKING}}: a valid pointer has its associated bit set which can be easily queried.

CHERI has been implemented on ARM64, RISC-V, and MIPS. Several hardware implementations are available, including Arm's Morello development board~\cite{arm_ltd_arm_2022}, SCI's CHERIoT RISC-V devices~\cite{SCI_Semiconductor}, and Codasip’s X730 RISC-V core~\cite{Codasip_2024}.

\section{Design of \bfsfork} \label{sec:design}

The overarching design goal of \sfork is to transparently support all of POSIX \fork's semantics within a (truly) single address space on modern hardware, without losing the lightweightness benefits of SASOSes.
Next, we set out design requirements and discuss the challenges that arise from them.
We then detail our approach to address these challenges.

\subsection{Design Requirements}

\paragraph{(R1) Lightweight \fork.} \label{R1}
The guiding benefit of SASOSes, lightweightness, must be maintained, ensuring high system performance, i.e. fast IPC and context/security domain switches (across processes and user/kernel), as well as a low per-process memory overhead.
This implies that \sfork must not re-introduce multiple address spaces which require costly hardware flushes and context switches, or require costly traps during context switches.
\sfork itself must also have a low latency (the time taken to \fork a process), 

\paragraph{(R2) \label{R2} Transparent implementation and compatibility.}
The differences between a strict POSIX \fork implementation and \sfork must not be visible to the application, i.e. using \sfork must not require application changes.
The system's behavior upon and following a call to \sfork (process state duplication) must also be equivalent to that of \fork from both the parent and child perspectives.

\paragraph{(R3) Strong isolation.} \label{R3}
\sfork must enforce the same level of isolation across \processes, and between \processes and the kernel, as enforced by POSIX processes: \processes cannot access the kernel/each other's memory, and can invoke the kernel only through system calls.

\paragraph{(R4) Flexibility to different SASOS use-cases.} \label{R4}
As discussed in \S\ref{subsec:background:sasoses}, SASOSes are a diverse class of OSes with different designs and goals.
To fit this diversity, \sfork should minimize the engineering effort required to implement its design in an existing OS.
Integrating \sfork should not compromise on the target OS key design aims e.g., it should be possible to disable isolation if a use-case does not require it.

\subsection{Design Approach and Challenges}
At a high level, \sfork emulates POSIX processes (\itprocesses), and upon a call to \fork, copies the memory used by the parent \process to a different location within a unique address space, for use by the child \process.
This approach, combined with \textbf{\hyperref[R1]{(R1) - (R4)}}, leads to two key challenges:

\paragraph{(C1) Memory Reference Relocation.} \label{C1}
In a traditional \fork implementation, which relies on creating a copy of the page table used by the calling process, the address space of the forked child is identical to that of its parent.
However, when forking in a single address space, the child's memory must reside at a different location in the same virtual address space as the parent.
We assume the use of position independent code, hence the vast majority of memory references will be relative to the stack, base, or instruction pointers.
However, to satisfy \textbf{\hyperref[R2]{(R2)}} and \textbf{\hyperref[R3]{(R3)}}, we must \emph{relocate absolute memory references}: these are memory references (pointers) located in the child's memory which still refer to parent memory after the copy.
They must be relocated to refer to child memory after the \fork, to ensure isolation and equivalent behavior.
Identifying such references at runtime (pointer tracking) is a hard problem~\cite{lu_how_2016}, as regular integers can be misidentified as memory references~\cite{wang_ramblr_2017}.

\paragraph{(C2) Process Isolation.} \label{C2}
POSIX processes are isolated by virtue of residing in different address spaces.
Thus, to satisfy \textbf{\hyperref[R3]{(R3)}}, we must isolate \processes within a single address space.
In turn, \textbf{\hyperref[R1]{(R1)}} and \textbf{\hyperref[R4]{(R4)}} require this isolation to be performed in a low-overhead and customizable manner.

\subsection{Trust \& Threat Models} \label{subsec:design:threatmodel}

\sfork assumes the same trust and threat model as POSIX \fork.
\sfork's Trusted Computing Base (TCB) is the OS kernel, which includes the implementation of \sfork.
The OS kernel distrusts all user code, encapsulated by one or more \sfork processes (\processes).
We assume that OS kernel interfaces properly sanitize user inputs.
Where \fork is used for isolation across \processes, we assume that attackers compromising one \process actively try to compromise the rest of the system's confidentiality, integrity or availability through shared memory, IPC, and kernel interfaces.
To that list we add another attack vector which is direct addressing, as \processes run within a single address space.
We assume that user code using the \fork API for isolation properly sanitizes cross-\process interfaces and appropriately sets system access-control permissions upon \fork since such applications were designed with this trust model in mind.
We consider side channels, including transient execution attacks, out of scope of this work.

\subsection{Design Overview}
\label{subsec:design:overview}

We propose a design based on six building blocks to address the requirements and challenges formalized earlier.
We elaborate on these in the subsequent subsections.

\begin{enumerate}[leftmargin=0.5cm,noitemsep]

\item \textbf{\bfprocesses.}
In a \sfork system, each thread is associated with a \process ID (PID). Each \process may have many threads.
Spawning a new \process creates a new thread with a new PID.
This matches the semantics of \fork, which copies a single thread \textbf{\hyperref[R2]{(R2)}}.

\item \textbf{Isolate \bfprocesses with CHERI.}
Satisfying \textbf{\hyperref[C2]{(C2)}} requires isolation within the address space.
This can be effectively addressed with CHERI's intra-address-space isolation~\cite{watson_cheri_2015, kressel_software_2023, esswood_cherios_2020}, where all memory references are bounded and so can be restricted to their own \process.

\item \textbf{Leverage CHERI tags to relocate memory references.}
Identifying absolute memory references to relocate in a child \process is required to satisfy \textbf{\hyperref[C1]{(C1)}}.
Memory tagging offered by CHERI~\cite{woodruff_cheri_2014} can be leveraged to reliably identify such references for relocation: upon \fork, memory references can be easily identified if a valid tag is set, and relocated to the child \process.

\item \textbf{Leverage Position-Independent-Code (PIC) to limit the number of references to be relocated.}
By compiling applications as PIC, most memory references are relative either to the stack/base pointer or to the program counter (PC).
They can thus be relocated without changes, ensuring the low \sfork latency required for lightweightness \textbf{\hyperref[R1]{(R1)}}.

\item \textbf{Revisit \CoW (CoW) for \bfprocesses.}
POSIX \fork implementations reduce the overhead of \fork by only copying pages upon writes.
This optimization cannot be applied as-is with \sfork, as the child accessing a page in read mode containing absolute memory references must trigger a copy of that page to perform the corresponding relocations, to prevent the child \process loading and using stale memory references which still point to the parent \process.
Hence, in its basic form \sfork requires copying each page accessed in write mode by the parent or the child, but also each page accessed in read mode by the child, something that we generalize as \CoA (CoA).
To maintain lightweightness \textbf{\hyperref[R1]{(R1)}}, CoA can be optimized into \CoPA (CoPA), which lets \processes share memory until a \process writes to it, or a child loads an absolute memory reference.

\item \textbf{Parameterized isolation.}
There are valid use-cases for different levels of isolation \textbf{\hyperref[R4]{(R4)}}: real-world systems may assume that processes distrust each other; that the entire system is trusted but contains bugs; or even that the entire system is trusted to function correctly.
We design \sfork to cater for any of these design points.
\end{enumerate}

\subsection{\bfsfork: Forking a \bfprocess}

\begin{figure}
\includegraphics[width=0.49\textwidth]{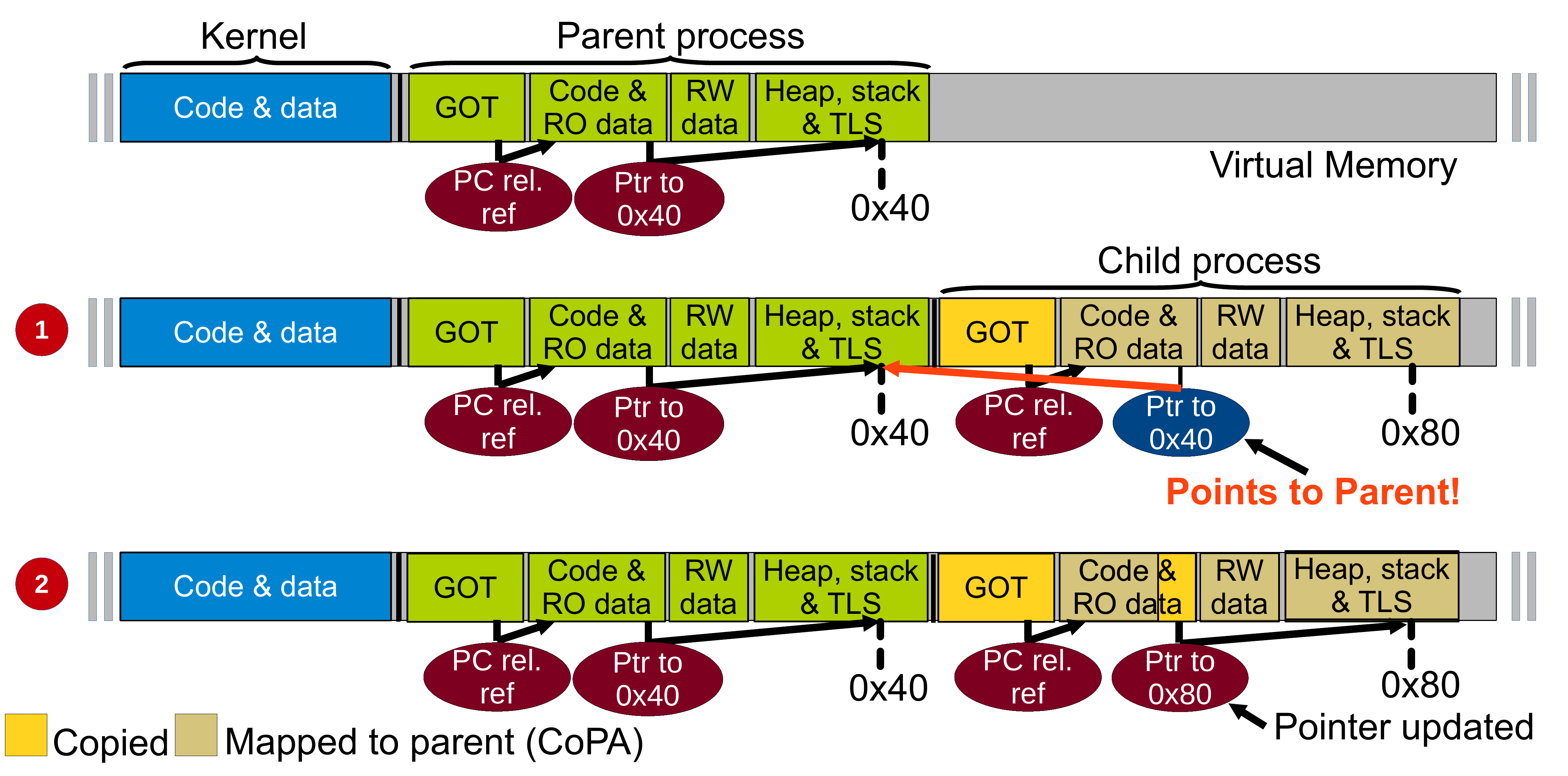}
\caption{Memory layout of \sfork. \BC{1} displays memory directly after a \fork. The child is mapped to the parent pages and absolute memory references point to parent memory. \BC{2} displays memory after the child accesses a page in the "code and read-only data" area containing an absolute memory reference: this page has been copied, and the memory reference relocated to point to child memory.}
\label{fig:sfork-overview}
\end{figure}

We now walk through the steps taken when a \process calls \fork to concretize this design.
This section is designed to be read alongside Figure~\ref{fig:sfork-overview}.

\paragraph{1. Parent State Duplication}

When a \process forks, \sfork reserves enough contiguous virtual memory to accommodate the entire child \process including the stack, heap and other data.
The parent page table entries are copied: initially, almost all pages of child memory are mapped to the same physical pages as the parent \process, similarly to a traditional CoW approach.
Any attempt to modify the heap, stack, TLS or other RW memory will result in a copy of the page written being made, along with relocation of absolute memory references should the page contain any.
As discussed, traditional CoW does not apply as-is, we thus rework it into CoPA to ensure that all absolute memory references contained in pages read by the child are relocated to their correct \process locations before use.
We discuss CoPA in detail in \S\ref{CoW}.

This mapping stage is illustrated by \BC{1} in Figure \ref{fig:sfork-overview}.
At this stage, pages have not yet been accessed: parent and child \processes share memory and memory references are not updated.
Pages containing memory-allocator metadata and global offset table (GOT) entries are proactively copied and updated during \fork to ensure that the new \process accesses the correct memory when loading memory references via the GOT or when performing heap operations.
In addition to copying the memory associated with a \process, relevant system resources are also duplicated as mandated by POSIX, \eg open file and message queue descriptors.

\paragraph{2. Post-Copy Phase}

At this stage, the memory and resources of the parent \process have been copied or mapped where appropriate for the child \process to use.
With the child \process memory and resources set up, \sfork generates a new \process ID (PID) and stores it in a memory location which cannot be modified by any \process.
The PID is used by the kernel to index per-process system resources such as file descriptor tables.
Following this, \sfork creates a new thread to run the \process.
The initial state of the child \process is the same as that of the parent, except that 1) the entry point is located within the code in the child \process's memory area following the call to \fork, and 2) any absolute memory references contained in registers are relocated to the proper locations in the child's memory (tags extend to values in registers, allowing differentiation of pointers from integers).
As the child \process accesses pages that contain references or writes to memory, pages are copied, as illustrated in \BC{2}.

\subsection{Isolation in \bfsfork} \label{subsec:design:isolation}

Isolation is required by \sfork at two levels: between \\ \processes (inter-\process), and between \processes and the kernel.
With regular POSIX \fork, inter-process isolation is enforced through having separate page tables for each \process.
This is not the case in \sfork, where isolation must be enforced with lightweight intra-address-space isolation.
\sfork also implies a trust boundary between \processes and the kernel.
Here, two main vectors of attack must be considered.
First, the system call interface must be hardened: the kernel must perform various validity checks on system call parameters, and also copy objects passed by reference to protect against time-of-check-to-time-of-use (TOCTTOU) attacks in concurrent systems~\cite{bhattacharyya_midas_2022}.
Second, for SASOSes such as unikernels which execute the kernel and applications at the same processor privilege level, \processes must be prevented from executing privileged instructions.

Still, an important point of our approach to isolation is to recognize that not all use-cases have the same needs for isolation and the underlying trust and threat models of the application must be preserved.
In cases where processes are assumed to distrust each other, e.g. when \fork is used for privilege separation, adversarial fault isolation is needed and full isolation is required.
An example of this is qmail~\cite{bernstein_qmail_2007} where processes are used to isolate components such as the SMTP server.
This corresponds to \textbf{\hyperref[U3]{U3}}.
However, in many cases the entire program is trusted, but may contain bugs that should ideally trigger faults when occurring, e.g. a bad reference outside of the process' accessible memory.
Web servers such as Nginx~\cite{noauthor_nginx_2024} use \fork (\textbf{\hyperref[U2]{U2}}) for increased throughput by leveraging concurrency, and use this fault protection trust model.
In such a case, production settings may make use of non-adversarial fault isolation, enabling memory isolation and simple kernel checks, but opting out of more expensive isolation primitives such as kernel-level TOCTTOU protections.
In other cases, the entire system may be trusted to function correctly, in which case production deployments may simply disable isolation.
An example is Redis using \fork for performing a backup snapshot through \cow: the child is simple and operating on trusted input, and thus is unlikely to trigger a bug and protections can be lifted.
We design \sfork to be flexible and accommodate for these different scenarios.

\subsection{Memory Layout of a \bfsfork-enabled SASOS}

In a system using \sfork, the kernel and application are compiled as PIC/PIE and each \process is loaded in a contiguous area of the virtual address space, as illustrated in Figure \ref{fig:sfork-overview}.
This allows intra-address-space isolation mechanisms relying on contiguous bounds to easily set and check these bounds, thus restricting \sfork \processes to their private memory.
Finally, a global offset table (GOT), providing a mechanism for PIC to locate global objects and functions in memory, must lie within the bounds of each \process.
In a system using a standard \fork implementation, the GOT is not modified since the address space layout does not change and memory references remain the same in the parent and child.
A child \process in \sfork is located at a different part of the address space, thus the GOT is copied and modified during the \fork to point to the correct locations within the new \process.
The application and kernel use disjoint areas for dynamically allocated memory.
ASLR can be implemented by randomizing the base offset of the contiguous memory area dedicated to each \process.
Supporting shared memory between \processes would be straightforward, requiring \texttt{shm\_open} to return a file descriptor representing an area of shared memory and then map the same set of physical pages within the virtual address space areas of relevant \processes.
Similarly, shared libraries can be supported by mapping those libraries in each \process when mapping a binary and creating capabilities with the proper permissions.

\subsection{Copying Memory: CoW vs. CoA vs. CoPA} \label{CoW}
\CoW (CoW) is a common optimization which reduces the overhead and latency associated with \fork~\cite{kulkarni_optimizing_2012,smith_effects_1988}.
When memory is marked as CoW, that memory is shared between processes.
As mentioned earlier, this is not achievable with \sfork due to the need for relocations within pages accessed in read (and write) mode and containing absolute memory references.
To mitigate this issue, in its basic form \sfork requires \CoA (CoA), where memory is initially shared by the parent and the child but marked as inaccessible, thus any access will trigger a copy and relocations.
We further propose \CoPA (CoPA), an optimization to CoA where memory can be shared in read-only mode like CoW, with an additional constraint that if the child loads memory references, the containing page is copied to relocate the references it contains first.
This means that non memory reference loads do not trigger copying.

\section{Implementation} \label{sec:implementation}

We implemented a prototype of \sfork on top of Unikraft~\cite{kuenzer_unikraft_2021}, a modern, actively maintained \sasos.
The core motivation for implementing our prototype with Unikraft is its large compatibility with unmodified applications~\cite{lefeuvre_loupe_2024} which will also run transparently with our implementation of \sfork, as well as its ability to run both bare-metal and on top of a hypervisor.
Further, Unikraft has been leveraged in several research projects~\cite{sartakov_cubicleos_2021, lefeuvre_flexos_2022, Duy2025} that explore the design space of SASOSes, making it particularly suitable to satisfy \textbf{\hyperref[R4]{(R4)}}.
Unikraft was also the basis of prior works on supporting \fork in SASOSes~\cite{lupu_nephele_2023}.

\paragraph{\bfsfork: Building Blocks}

\sfork requires two fundamental building blocks: absolute memory reference identification (\textit{\textbf{\hyperref[C1]{C1}}}) and an intra-address-space isolation mechanism (\textit{\textbf{\hyperref[C2]{C2}}}).
Our prototype implementation uses the CHERI hardware capability model~\cite{woodruff_cheri_2014, watson_cheri_2015} that offers all the above-mentioned features.
CHERI supports memory tagging and also offers intra-address-space memory protection by ensuring that every memory reference is checked against the bounds of the object it refers to.
CHERI uses a sealing mechanism~\cite{watson_capability_2023} to enable exception-less security domain switches: we leverage it for user/kernel isolation, satisfying \textbf{\hyperref[R1]{(R1)}}.
Sealed capabilities are data structures available to \processes which, when used, trigger a safe transition to a predetermined and unforgeable location -- in our case the system call handler in the kernel.
Finally, CHERI offers a \textit{fault on capability load}~\cite{MorelloISA, CornucopiaReloaded} feature, with which we implement CoPA.

\paragraph{Implementation Overview}

We ported Unikraft to CHERI on the ARM Morello platform~\cite{arm_ltd_arm_2022} (\S\ref{subsec:impl:port-to-cheri}), and to execute on top of the bhyve hypervisor on CheriBSD~\cite{watson_cheri_2015} (CHERI-enabled FreeBSD) to leverage VirtIO networking on ARM Morello.
We implemented \sfork as a new kernel library in Unikraft.
In total, we changed 7~KLoC in the kernel: 4,230~LoC for CHERI and bhyve support in Unikraft, and 3,000~LoC for \sfork.
Applications which run on Unikraft do not require porting to work with \sfork.

\subsection{Porting Unikraft to CHERI} \label{subsec:impl:port-to-cheri}

Porting Unikraft to CHERI (in ``pure-capability mode''~\cite{watson_capability_2023}) required a modest engineering effort.
Key changes needed were 1) in the early boot code to initialize CPU capability features and registers (e.g. exception vectors); 2) in the virtual memory allocator (tinyalloc~\cite{noauthor_thi-ngtinyalloc_nodate}) to comply with CHERI's 16-byte pointer alignment requirements and set bounds on allocated memory; and 3) in broader kernel code to replace pointer arithmetic with explicit pointer operations as capabilities are not interchangeable with integers~\cite{CHERI_GUIDE}.

\subsection{Implementation of \bfsfork}

\paragraph{Assigning capabilities to \bfprocesses}

The key security invariant the kernel must enforce to maintain isolation \textbf{\hyperref[R3]{(R3)}} is that all capabilities (pointers) available to a \process only grant access to memory falling within the area of the virtual address space allocated to this \process.
At boot, \sfork initializes capabilities for the kernel and applications.
The sealed capabilities which are needed for trapless system calls are also set up at this stage.
The above-mentioned security invariant is enforced upon requests from the application for memory allocation (e.g. \texttt{mmap}/\texttt{brk}), as well as upon and after \process creation through \sfork, where all absolute memory references must be relocated to avoid leaking capabilities from the parent to the child.
This includes among others the program counter capability (PCC), whose bounds and permissions are used in PIC (see \S\ref{subsec:design:overview}) for relative references such as function calls (\cf Figure \ref{fig:sfork-overview}).

Our \sfork prototype follows the memory layout of Figure \ref{fig:sfork-overview}.
Each \process owns a private, statically-allocated heap with a build-time-configurable size, from which the majority of dynamic memory allocation requests are served.
This is done to reduce complexity in the TCB but can be replaced with dynamic heaps.
The kernel ensures anonymous \texttt{mmap} requests are served by returning capabilities pointing to the calling \process virtual memory area.
Our modular prototype \textbf{\hyperref[R4]{(R4)}} enables different trade-offs, \eg using a single shared isolation-aware heap~\cite{Amar2025}, or per-\process heaps~\cite{lefeuvre_flexos_2022}.

\paragraph{\CoPA} \label{CoPA}

\begin{figure}
\includegraphics[width=0.47\textwidth]{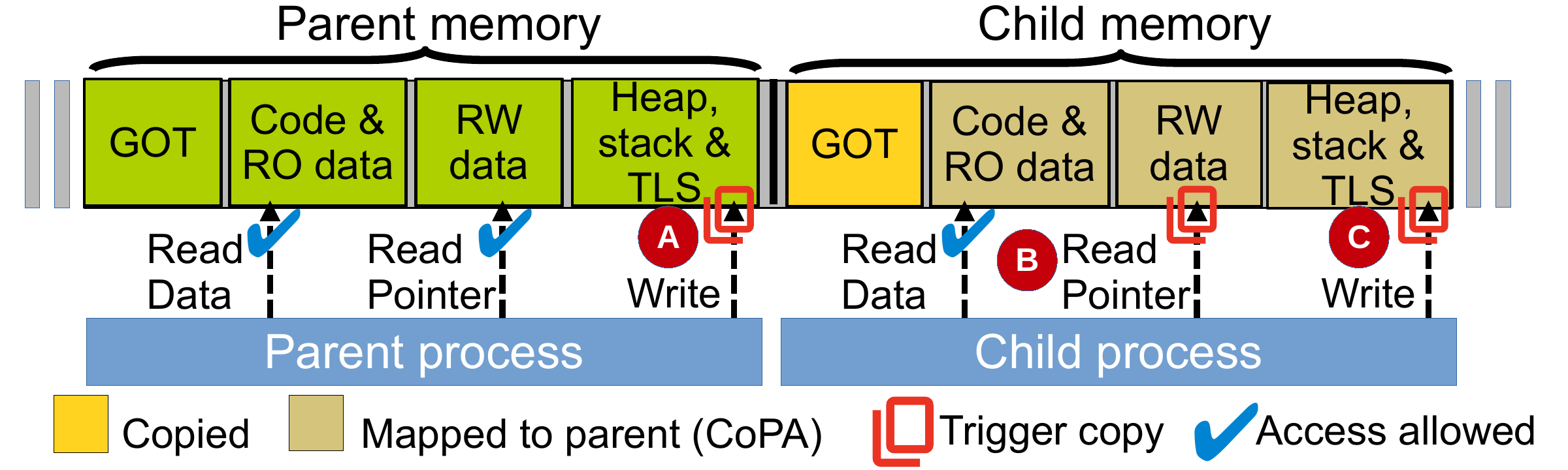}
\caption{\CoPA. Writing to pages (\BC{A} and \BC{C}), or the child loading from pages that contain pointers (\BC{B}), triggers copying.}
\label{fig:copa}
\end{figure}

We implement CoPA using an additional page-table permission bit present with CHERI, which triggers a fault when a capability is loaded from that page.
Loading references from the parent \process and read accesses do not trigger copies.

With this approach, following a \fork, page copies are triggered by writes from either the parent or the child, or capability loads by the child.
The copy follows three steps.
First, the child page table entry is changed to point to a free physical page in memory and remains inaccessible until the copying is finished.
Then, the page is copied.
Finally, the copied page is scanned in 16-byte increments, which corresponds to the size of a CHERI capability, for absolute memory references pointing to parent memory: references are identified by the presence of a valid CHERI tag.
For each reference with a valid tag, its target and memory bounds are checked: if it points outside of the \process's dedicated memory area in the virtual address space or has bounds which allow access outside the memory area, the reference is relocated to the correct location in the child \process with bounds which are restricted to that \process only.
Thus, all references are found and updated to point into child \process memory.
Figure \ref{fig:copa} shows CoPA in operation: the child attempts to write to a page of memory (\BC{C}) and read a pointer from an implicitly shared page (\BC{B}), resulting in a transparent copy similarly to standard CoW.
Similarly, the parent attempts to write to a page of memory (\BC{A}), triggering a copy.
This approach covers all possible references; absolute references are identified via a valid tag, and all references derived from these are tightly bounded to the \process since CHERI bounds cannot increase; relative references use a base absolute reference (e.g. stack/base) which are identified and updated.

\subsection{Cross-\bfprocess Isolation}

Inter-\process isolation with CHERI is enforced following two principles.
First, \emph{\itprocesses cannot increase their privileges.}
By design, the privileges associated with a CHERI capability (e.g. bounds of the area it points to) cannot be increased (\emph{monotonicity})~\cite{watson_capability_2023}.
A \process thus cannot forge a capability with greater privileges to escape its protection domain.
Second, \emph{capabilities do not leak across \itprocesses.}
Memory copies are triggered by CoPA in all cases that may enable a child \process to access a stale capability.
As part of the copy, parent capabilities are relocated and thus not exposed to child \processes.
Integers can be used for relative addressing, however, these are subject to the bounds of the capability they are relative to, such as the program counter, all of which are restricted to the local \process.

\subsection{\bfprocess-Kernel Isolation} \label{subsec:implementation:kernel-isolation}

User-kernel isolation follows 4 principles.
First, \emph{kernel entry-points are protected.}
In our prototype, system calls are performed through safe transitions using sealed kernel code capabilities~\cite{watson_capability_2023}.
Sealed capabilities restrict kernel entry points and there is no other way for a \process to invoke kernel code.
Such restriction is similar to that of traditional system call instructions, however it is achieved without the need for a costly trap, to comply with \textbf{\hyperref[R1]{(R1)}}.

Second, \emph{\itprocesses cannot execute privileged instructions.}
Our \sfork prototype executes the \processes and kernel at the same processor privilege level (EL1).
To prevent user code from executing system instructions which could compromise the security of the system, we leverage the CHERI system permission capability permission bit~\cite{watson_capability_2023}.
\process capabilities do not have this permission and thus cannot execute privileged instructions including \texttt{MSR} and \texttt{MRS} instructions.
This approach satisfies \textbf{\hyperref[R1]{(R1)}} through avoiding the need for costly memory scanning~\cite{dautenhahn_nested_2015}.

Third, \emph{data flowing into the kernel is validated.}
A malicious \process may attempt to pass corrupted data to the kernel through system call arguments.
This is a major source of security vulnerabilities in commodity OSes~\cite{wang_check_2018, yamaguchi_chucky_2013, li_path-sensitive_2022, bagherzadeh_analyzing_2018, li_arbitrar_2021}, which may enable attackers to e.g. trick the kernel into leaking data or capabilities to another \process' memory.
Carefully sanitizing untrusted inputs becomes necessary in much the same way as other OSes.
Still, not all deployments of \sfork may have an adversarial threat model \textbf{(R4)} (\S\ref{subsec:design:isolation}), thus we make it easy to disable checks for a given deployment.

Fourth, \emph{\itprocess buffers passed by reference to the kernel are copied.}
TOCTTOU vulnerabilities are possible if a kernel input can be modified by a \process between a check and its use, negating the check~\cite{bhattacharyya_midas_2022, xu_precise_2018, xu_warpattack_2023}.
To prevent this we copy values to kernel memory before checking and upon completion, back to user memory, a mechanism used in commodity OSes.
Following our flexibility design goal \textbf{\hyperref[R4]{(R4)}}, we make it possible to disable TOCTTOU protections for non-adversarial threat models.

\subsection{Retrofitting \bfsfork into an existing SASOS}

Many implementations of \sfork will be \emph{retrofitted} into existing SASOSes, rather than built into entirely new operating systems.
The aforementioned address space duplication approach of \sfork is OS-agnostic, however introducing multiprocessing in a SASOS will also require per-OS engineering efforts.
We describe such effort here, showing that for our prototype based on Unikraft it was relatively modest.

\paragraph{Per-Process Kernel State.}
SASOSes which are designed to run a single process, such as unikernels~\cite{kuenzer_unikraft_2021, madhavapeddy_unikernels_2013, lefeuvre_flexos_2022, kressel_software_2023, olivier_binary-compatible_2019}, may not support per-\process state: file descriptor tables, task structs, PIDs, process scheduling, per process signals, among others.
This is because assuming that there is no more than one process in the system allows simplifying the development of these OSes and of the POSIX-like compatibility layers they generally expose~\cite{lefeuvre_loupe_2024}.
The amount of refactoring required to support per-process kernel state will be variable on a case-by-case basis, depending on the target SASOS, and can be non-negligible.
For our prototype we added the OS process features needed for the applications presented in evaluation including per-\process file descriptors and task structs, PIDs and scheduling.
Further, we refactored the implementation of system calls such as \texttt{wait} and \texttt{getpid} to facilitate \process support, and added support for per-\process heaps.

\paragraph{User-Kernel Separation.}
SASOSes such as unikernels are designed with no user/kernel isolation.
This does not allow for support of the generic threat model of POSIX \fork (\S\ref{subsec:design:threatmodel}), unless additional protections are implemented (\S\ref{subsec:implementation:kernel-isolation}).
Still, the absence of isolation may be acceptable in the threat model of such OSes \textbf{\hyperref[R4]{(R4)}}.
Unikernels may for example want to implement \fork for concurrency or \cow snapshot patterns, but not for privilege-separation patterns due to their design choices.
In Unikraft, we implemented the checks discussed in \S\ref{subsec:implementation:kernel-isolation} to showcase the cost of different levels of protection.
We also extended Unikraft's build system to differentiate between application and kernel code through linker scripts, which helps to locate the kernel and applications in contiguous areas of the virtual address space.

\paragraph{Symmetric Multiprocessing (SMP) Support.} \label{item:implementation:smp}
SASOSes designed for a single process may not support SMP, an important feature required for certain \fork use cases, e.g. to exploit concurrency.
Hence, it may be necessary to retrofit SMP into the OS, which comes too at a non-negligible cost.
Unikraft currently supports SMP with a ``big kernel lock'', letting application code run concurrently but serializing kernel code execution.
Optimized support is under active development by the community.

\section{Evaluation} \label{sec:evaluation}

The evaluation aims to answer the following research questions (RQs):

\vspace{0.1cm}

\noindent \textbf{RQ1}: \textit{How does \sfork compare to a monolithic kernel and existing SASOS \fork implementations on key lightweightness metrics?}
We use Redis, MicroPython and microbenchmarks to examine \fork latency, memory consumption, and IPC performance.

\noindent \textbf{RQ2}: \textit{How does the overall performance of applications running on \sfork compare to a monolithic kernel's \fork?}
We examine overall system performance through popular \fork-based applications: Redis, MicroPython, and Nginx, compared to CheriBSD.

\noindent \textbf{RQ3}: \textit{Does \sfork enable applications to unlock the benefits (e.g. concurrency) of \fork effectively?}
We evaluate concurrency with MicroPython and Nginx compared to CheriBSD.
Additionally, Redis leverages CoW to enable its background database dump feature.

\noindent \textbf{RQ4}: \textit{How does CoPA compare to CoA and a full copy?}
We use Redis to examine the effect of CoPA compared to CoA and synchronously copying memory upon \fork.

\vspace{0.1cm}

We compare \sfork with CheriBSD~\cite{CHERIBSD_WEBSITE}, a CHERI port of FreeBSD, and Nephele~\cite{lupu_nephele_2023}, a virtua\-lization-based approach to support \fork in a unikernel.
Cheri\-BSD is the most mature capability-aware monolithic kernel running on the Morello platform, and Nephele represents the most-recent attempt at supporting \fork in a SASOS (by re-introducing multiple address spaces).
We do not include a non-CHERI baseline since this introduces additional variables; the CHERI/Morello prototype hardware and software stack is still not fully mature and thus software running in pure-capability mode faces in some situations non-negligible overheads that have been well-docu\-mented~\cite{MORELLO_MICROPYTHON}.
Analysis has shown that the majority of these overheads can be eliminated in future hardware implementations~\cite{watson2023early}, reducing the overhead to a negligible level (1.8 - 3\%).

All comparisons are made between \sfork executing on top of the bhyve hypervisor on CheriBSD 23.11, and CheriBSD 23.11 running natively in pure capability mode and with performance optimizations enabled~\cite{noauthor_benchmarking_nodate}.
\sfork needs to run virtualized as the Unikraft kernel we build upon lacks I/O drivers beside VirtIO.
We encountered difficulties running CheriBSD virtualized, and note that running it bare metal gives it a conservative advantage vs. \sfork executing on top of bhyve.
All experiments are run on the ARM Morello development system~\cite{arm_ltd_arm_2022} with 4 ARMv8.2-A cores @ 2.5 GHz and 16 GB RAM.
Nephele~\cite{lupu_nephele_2023} is designed for x86\_64 only.
For this reason direct performance comparisons between applications running on \sfork and Nephele cannot be achieved.
For the sake of completeness, we replay Nephele's relevant microbenchmarks (fork latency, memory consumption) with \sfork and compare the results with numbers extracted from Nephele's paper.
Further, we examine the effect of TOCTTOU protections and how our CoPA optimization compares to both CoA and a full synchronous copy of the parent's memory upon \fork.
The graphs report averages of 10 measurements, with standard deviation as error bars (in many cases hardly visible because of the results' stability).

\subsection{Real-World \fork-based Use-Cases}
We examine \sfork through three popular \fork-based applications, Redis~\cite{noauthor_redis_2024}, MicroPython~\cite{noauthor_micropython_2024, noauthor_micropython_nodate, MORELLO_MICROPYTHON}, and Nginx~\cite{noauthor_nginx_2024}.
Each application utilizes \fork for a different purpose, and is impacted differently by the characteristics of this primitive.

Redis uses \fork and its on-demand CoW state duplication capabilities to create a database snapshot and save it to the disk in the background (\textbf{\hyperref[U2]{U2}} + \textbf{\hyperref[U4]{U4}}). Here, \fork latency is important, since the main database cannot handle requests during this time.
We use Redis for \textbf{RQ1}, \textbf{RQ2} and \textbf{RQ3}.

MicroPython is a lightweight implementation of Python, a language runtime commonly used in Function as a Service (FaaS) frameworks~\cite{shafiei_serverless_2022}.
FaaS relies on low cold start times to rapidly handle incoming requests and low resource consumption to handle as many requests as possible~\cite{shen_defuse_2021, kanas_diminishing_2024, liu_faaslight_2023}.
\fork is a compelling mechanism to achieve this~\cite{ao_faasnap_2022, li_serverless_2022, oakes_sock_2018}, by spawning new instances from an already initialized language runtime (\textbf{\hyperref[U2]{U2}} + \textbf{\hyperref[U5]{U5}}).
These requirements map directly to low \fork latency and memory consumption (\textbf{RQ1}).
In addition, exploiting concurrency, as enabled by \fork, is crucial in FaaS due to the stateless nature of functions (\textbf{RQ3}).

Finally, Nginx uses \fork to spawn worker processes, isolated from each other, and handling requests concurrently (\textbf{\hyperref[U5]{U5}}), used to answer \textbf{RQ3}.
Unlike the other applications evaluated, here \fork latency does not play a significant role, since Nginx workers are long-lived.

\paragraph{Redis snapshots}

\begin{figure}
\includegraphics[width=0.48\textwidth]{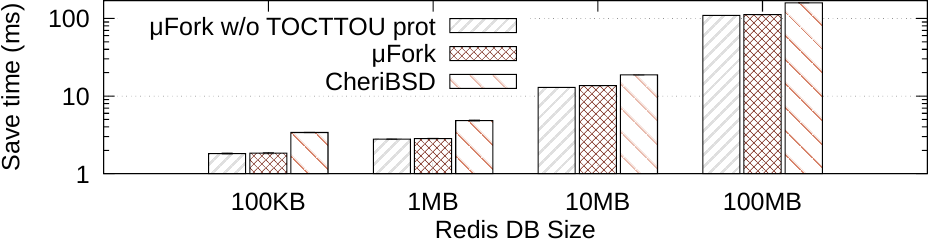}
\caption{Redis DB overall save times (ms).}
\label{fig:redis_save}
\end{figure}

We run Redis on CheriBSD and \sfork.
This experiment populates a Redis database with different amounts of 100~KB entries and then triggers a background save operation.
Both systems store the resulting database dump to a ram-disk, minimizing I/O latency.
Figure~\ref{fig:redis_save} shows the overall save times of Redis with database sizes ranging from 100~KB to 100~MB.
Across the range, \sfork outperforms CheriBSD.
\sfork is 1.9$\times$ faster than CheriBSD at 100~KB (1.8 vs. 3.4~ms), and 1.4$\times$ faster at 100~MB (109 vs. 158~ms).

\begin{figure}
    \includegraphics[width=0.48\textwidth]{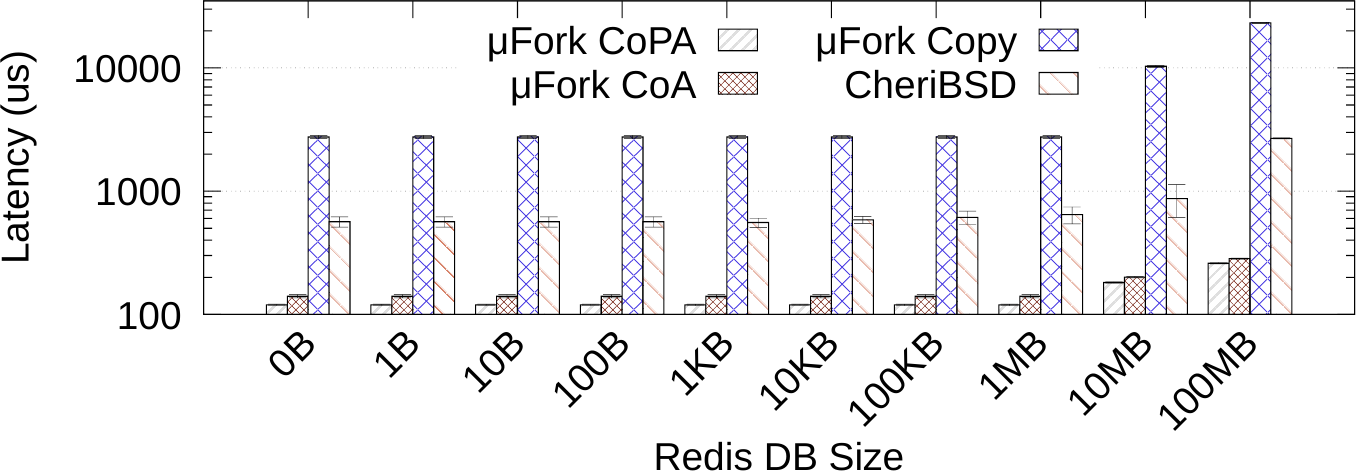}
    \caption{Redis \fork latency (us).}
    \label{fig:redis_latency}
\end{figure}

Figure~\ref{fig:redis_latency} shows the \fork latency (time needed for the \fork call to complete) caused by forking the main Redis process with various database sizes.
\sfork is consistently faster and more lightweight than an equivalent process forking on CheriBSD, by a factor of 5-10$\times$.
This latency difference can explain the larger difference in save times seen at smaller database sizes, although with larger databases \fork's latency is not the bottleneck in the overall database save operation.
Here, \sfork's lightweight state transfer method and page-table updates, as well as the lower cost of making exception-less system calls at the same exception level~\cite{kuenzer_unikraft_2021}, help to explain the difference.
The results also show that CoPA can reduce \fork latency by up to 89$\times$ vs. realizing a synchronous copy, and up to 1.18$\times$ vs. CoA since less memory must be copied.
The cost of TOCTTOU protection is relatively minor (2.6\% at 100~MB).

\begin{figure}
    \includegraphics[width=0.48\textwidth]{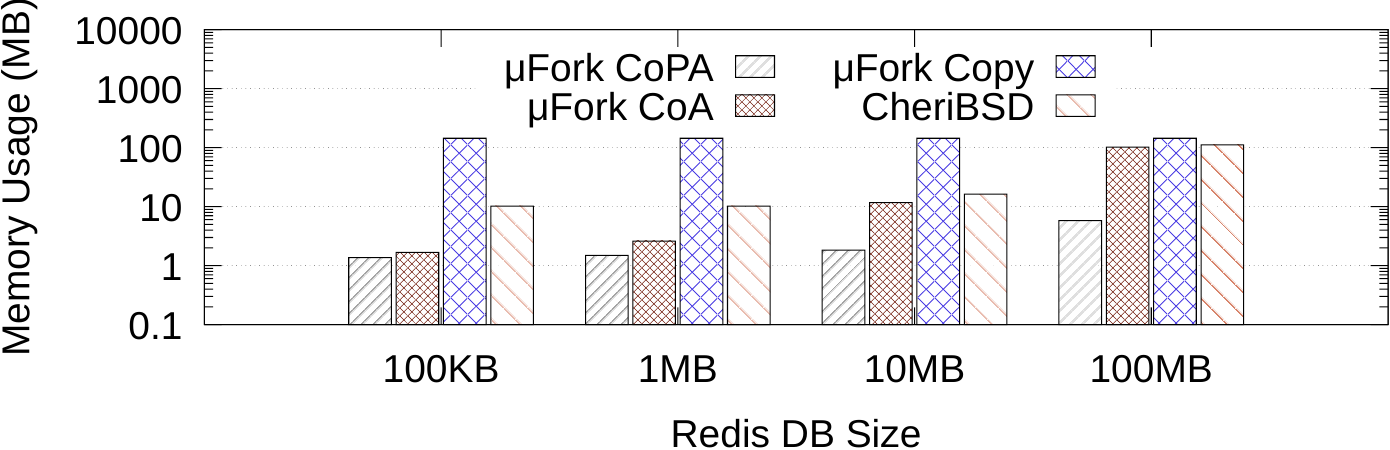}
    \caption{Redis memory consumption (MB).}
    \label{fig:redis_mem}
\end{figure}

The memory consumption of the forked Redis process with different database sizes also gives a real world example of how the lightweightness of \sfork translates to a real application.
Figure~\ref{fig:redis_mem} shows that the memory consumed by a forked Redis process on \sfork is significantly lower than on CheriBSD, 6~MB compared to 56~MB with a 100~MB database which can be explained by higher allocator memory consumption.
For context, at 100~MB database size, a forked Redis process consumes 7~MB on a standard aarch64 Linux install on the ARM Morello system meaning that the high CheriBSD figure is likely something which can be reduced with further optimization.

\paragraph{Function as a Service}

FaaS functions are typically short-lived, with 50\% of functions taking less than 1s to execute~\cite{shahrad_serverless_2020}.
Thus, FaaS aims to keep the time needed to start a new instance of a function before executing it (cold start) low, as long cold starts severely reduce the efficiency of the system.
This experiment uses the Zygote language runtime pre-warming technique~\cite{ao_faasnap_2022, shin_fireworks_2022, oakes_sock_2018}: the runtime is initialized (e.g. loading Python imports) once in a Zygote process, and subsequently each request is served by forking the Zygote into children that execute the function.
We used the \texttt{float\_operation} benchmark from FunctionBench~\cite{FUNCTIONBENCH}: to reduce the effect of I/O and system calls, it performs a series of calculations before returning. 
The experiment setup sees a single thread forking as many times as possible within a 10-second time window on both CheriBSD and \sfork.
Here, both systems are able to take advantage of concurrency over multiple cores.
The Morello CPU has 4 cores, 1 is used for the coordinating thread, and the rest for function execution.

\begin{figure}
    \includegraphics[width=0.48\textwidth]{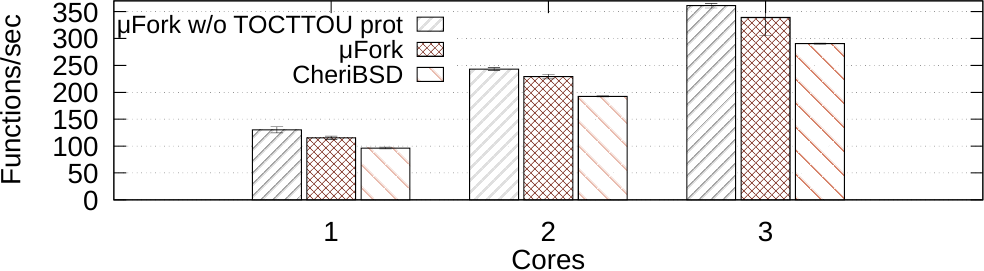}
    \caption{FaaS function throughput.}
    \label{fig:faas}
\end{figure}

Figure~\ref{fig:faas} shows the number of functions executed per second on 1 to 3 cores.
The benchmark does not perform system calls and I/O, thus function throughput is primarily impacted by \fork latency.
As a result, \sfork outperforms CheriBSD processes: its lower \sfork latency allows it to handle 24\% more requests.
The cost of TOCTTOU protection is negligible since the experiment is not system-call intensive.

\paragraph{Nginx multi-worker deployments}

\begin{figure}
\includegraphics[width=0.48\textwidth]{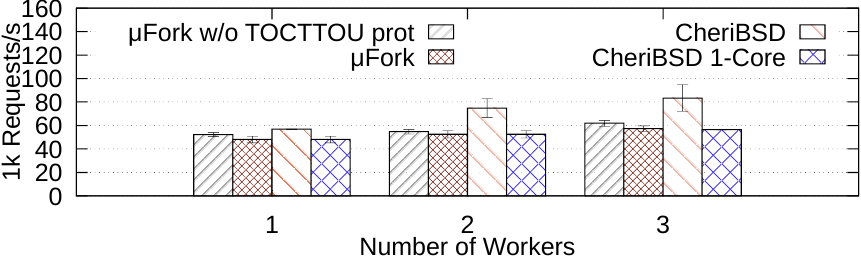}
\caption{Nginx throughput.}
\label{fig:nginx}
\end{figure}

Nginx forks workers to leverage concurrency and thus handle more requests.
For this experiment we use the \texttt{wrk} benchmark to measure the throughput of Nginx with 1, 2, and 3 workers on CheriBSD and \sfork.
Here, \sfork is limited by the immature support of SMP in Unikraft which hampers network performance across multiple cores (\S\ref{item:implementation:smp}).
This is not a limitation of \sfork, and the issue is currently being addressed by the Unikraft community.
For this reason, we were unable to get stable numbers on more than one core and Figure~\ref{fig:nginx} shows the performance of Nginx running on a single core only with \sfork.
Still we include these results to demonstrate compatibility: Nginx can run using \sfork.
Despite the single core restriction, we see a 15.6\% performance improvement by increasing the worker count on \sfork from 1 to 3, likely due to workers yielding during I/O.
We show CheriBSD performance both when allowed to scale to multiple cores and when restricted to a single core.
Unsurprisingly, CheriBSD outperforms \sfork across multiple cores, however, restricted to a single core, \sfork is able to handle 9\% more requests. 
The throughput cost of TOCTTOU protection is 6.5\% on average.

\subsection{Microbenchmarks}

\paragraph{\fork Latency}

\begin{figure}
    \includegraphics[width=0.48\textwidth]{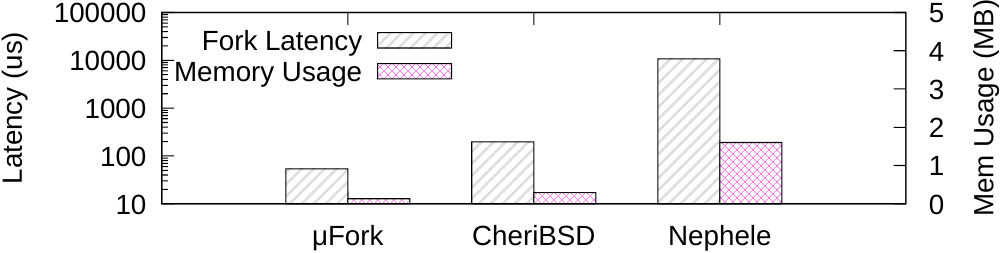}
    \caption{\fork latency and memory usage (hello world).}
    \label{fig:single_fork}
\end{figure}

\begin{figure}
    \includegraphics[width=0.48\textwidth]{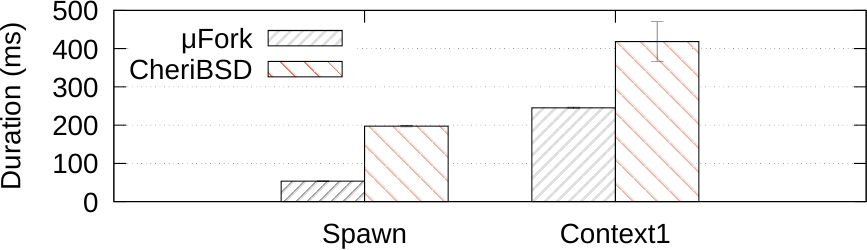}
    \caption{Unixbench \fork latency (Spawn) and IPC (Context1) benchmarks execution time.}
    \label{fig:unixbench}
\end{figure}

We measured the latency observed when forking a single minimal process (hello world C program), and using Unixbench Spawn~\cite{lucas_unixbench_2024} measuring the time taken to fork 1000 processes as fast as possible.
The results are presented on Figure~\ref{fig:single_fork} (hello world) and Figure~\ref{fig:unixbench} (Unixbench).
\sfork has a consistently lower latency vs. CheriBSD and Nephele.
Unlike Nephele, \sfork does not need to create a new Xen domain, which is notably a costly endeavor.
As a result, \sfork's \fork latency is several orders of magnitude smaller: 54~$\mu$s vs. 10.7~ms.
Similarly, \sfork outperforms CheriBSD, with a \fork latency of 54~$\mu$s vs. 197~$\mu$s, aided by its simple state transfer mechanism.
Looking at Unixbench Spawn, 1000 consecutive forks and exits are completed in 56~ms on \sfork and 198~ms on CheriBSD, attributable to the lower fork latency and the exception-less, single-privilege-level context switches between user and kernel mode in \sfork~\cite{madhavapeddy_unikernels_2013}.

\paragraph{Memory Consumption}

An important factor in considering how many processes a system can handle is memory consumption.
We consider the proportional resident set as the memory consumed by a process.
Figure~\ref{fig:single_fork} shows the memory consumed per process when forking a minimal application.
Again, \sfork shows its lightweightness, occupying only 0.13~MB compared to 1.6~MB for Nephele and 0.29~MB for CheriBSD.
This size difference can be attributed to memory consumed by shared libraries and the memory allocator.

\paragraph{Inter-Process Communication}
Fast IPCs are one of the main benefits of SASOSes, and \sfork unlocks them for the first time in \fork-based applications.
We use the Unixbench Context1 benchmark, which opens a pipe between two processes and increments a number until a set value is reached, in this case 100k.
Figure~\ref{fig:unixbench} shows that \sfork completes this in 245~ms compared to 419~ms on CheriBSD.
This difference in performance can be attributed to the exception-less, single-privilege-level system calls in \sfork.

\paragraph{CoPA \vs CoA \vs Full Copy}
The effect of CoPA versus CoA and upfront copy is shown in Figure~\ref{fig:redis_latency} and Figure~\ref{fig:redis_mem}.
As \sfork is built using large static heaps, the memory transferred by a full copy is correspondingly large, e.g. consuming 144~MB with a 100~MB Redis database, of which 136.7~MB is the large static heap.
So too is the \fork latency, taking in that case 23.2~ms.
CoA is cheaper, consuming 101~MB and thus \fork latency is lower at 283~$\mu$s since not all memory will be accessed by \processes.
This demonstrates that performance is still improved over an upfront copy on systems which cannot support CoPA.
CoPA reduces the consumed memory to 6~MB, and the \fork latency further to 260~$\mu$s with a 100~MB database because read accesses to shared memory are allowed by the child and parent.

\section{Discussion}

\paragraph{\textit{Alternatives to CHERI}}
Absolute memory references must be accurately identified to enable their relocation to another part of the address space to implement \sfork.
Beyond CHERI, this can be achieved with other hardware memory tagging technologies~\cite{bannister_mte_2019, serebryany2018memory, LOWRISC_TAGS, HDFI} that associate with each memory location a tag able to hold metadata regarding that location.
This tag could be used to mark memory locations containing absolute memory references.
That would need to be associated with compiler-generated instrumentation to propagate tags e.g. when a reference is copied or derived from another reference (something realized transparently by the hardware with CHERI).
Alternative software approaches can also be used, such as shadow memory~\cite{serebryany2012addresssanitizer}, maintaining a map of memory locations containing references, also set up and maintained with compiler-generated instrumentation.

\sfork further requires an intra-address-space isolation mechanism to isolate \processes.
Various approaches have been proposed to achieve this~\cite{Lefeuvre2025}: on the hardware side different flavors of memory protection keys~\cite{intel_mpk_2022, DONKY, ARM_MPK} provide intra-address-space page-level memory protection as well as exception-less (i.e. fast) protection domain switching, which is important for lightweightness.
Other hardware technologies such as Mondrian memory protection~\cite{witchel2002mondrian} offer byte-level isolation.
Software approaches such as software fault isolation~\cite{wahbe1993efficient} or fat pointers~\cite{necula2002ccured} may also be used, although the associated performance hit may be deterring.

Beyond these two main requirements that are necessary for a functional implementation of \sfork, an optional feature that is needed for our CoPA optimization is the capacity to fault on a subset of memory accesses: the loads of absolute memory references.
We are not aware of a mechanism other than CHERI offering the ability to fault on memory reference loads needed to implement CoPA.
However, as we show in the evaluation, systems which cannot support CoPA still show significant performance improvements from CoA.

\paragraph{\textit{Fragmentation}}
A limitation of \sfork is the fact that \processes are loaded into potentially large contiguous areas of virtual memory, raising concerns about fragmentation.
The 64-bit virtual address space we target is very large, and we expect that the vast majority of multiprocess applications will not exhaust that space.
For scenarios where fragmentation could be a problem (long-running applications, forking a high number of \processes requiring large contiguous amounts of memory), solutions including compacting the virtual address space periodically or using size classes akin to size-class memory allocators~\cite{silvestro_freeguard_2017, silvestro_guarder_2018, leijen_mimalloc_2019}, can be explored in future work.

\section{Related Work}
\paragraph{\textbf{\textit{\SASOSes}}}
SASOSes have been an active research topic since their inception in the 1990s.
Foundational works such as Opal~\cite{chase_sharing_1994}, Nemesis~\cite{roscoe_linkage_1994}, Angel~\cite{wilkinson_compiling_1993}, and Mungi~\cite{heiser_mungi_1998} simplified and improved the performance of IPCs and I/O by placing all processes and data in a single address space.
Since then, many types of SASOSes have been proposed, including single-level-store OSes~\cite{levy2014capability}, dataplane OSes~\cite{ix_dataplane}, software-isolated OSes~\cite{hunt_singularity_2007, boos_theseus_2020}, and unikernels~\cite{kuenzer_unikraft_2021, madhavapeddy_unikernels_2013, lefeuvre_flexos_2022, kressel_software_2023, sartakov_cubicleos_2021, olivier_binary-compatible_2019}.
The lack of \fork in SASOSes is a well-known issue that attracted significant research efforts~\cite{wilkinson_compiling_1993, heiser_mungi_1998, lupu_nephele_2023, zhang_kylinx_2018, tsai_cooperation_2014, li_iso-unik_2020}.
Other relevant SASOS research focuses on isolating components in a single address space~\cite{olivier_case_2020, sung_intra-unikernel_2020, esswood_cherios_2020, sartakov_cubicleos_2021, almatary_compartos_2022, almatary_case_2024, xia_cherirtos_2018}.
Notable examples include Singularity~\cite{hunt_singularity_2007} which isolates between processes using statically verified invariants, FlexOS~\cite{lefeuvre_flexos_2022} which uses build-time user-defined schemas to isolate components using different mechanisms~\cite{kressel_software_2023}, and RustyHermit~\cite{sung_intra-unikernel_2020} which uses Intel MPK for lightweight intra-address-space isolation.

\paragraph{\textit{Single-Address-Space \fork}}

As the historical process-creation primitive, \fork remains a widespread part of modern UNIX OSes.
Attempts have been made to support \fork within a single address space.
A first category of works, including Angel~\cite{wilkinson_compiling_1993} and Mungi~\cite{heiser_mungi_1998}, used segment-relative addressing to relocate the process.
As discussed earlier (\S~\ref{sec:segmentation}), this approach is complex and requires significant compiler and application porting work.
A more recent class of works sees the OS as a process and relies on an external entity to perform the fork: Graphene/Gramine~\cite{tsai_cooperation_2014, tsai_graphene-sgx_2017} and Vessel~\cite{Vessel_2024} run as a process and piggyback onto the host OS' \fork; KylinX~\cite{zhang_kylinx_2018} and Nephele~\cite{lupu_nephele_2023} rely on a hypervisor implementation of \fork.
Unlike \sfork these approaches reintroduce multiple address spaces and thus defeat the principle of SASOSes and rely on more expensive \fork implementations (host OS or hypervisor) which hurts overall performance.
Iso-Unik~\cite{li_iso-unik_2020} also re-introduces multiple address spaces within the SASOS, which hurts light\-weightness.
A last class of works focuses exclusively on \texttt{vfork + exec} patterns.
CHERI co-processes~\cite{noauthor_ctsrd-chericomsg_2024} support multiple processes (\vfork + \texttt{exec}) within a single address space to enable faster IPC.
Junction~\cite{Junction_2024} supports multiple instances of a process within an address space although without isolation.
Occlum~\cite{shen_occlum_2020} supports \texttt{posix\_spawn} within a single enclave.
However, this does not support the state duplication feature of \fork, and requires applications to be modified to use \texttt{posix\_spawn} instead of \fork.

\paragraph{\textit{Memory Reference Identification.}}
CARAT~\cite{CARAT} and CA\-RAT CAKE~\cite{CARAT_CAKE} replace hardware paging with a software implementation.
In doing so they encounter the same challenges of memory reference identification and intra-address-space isolation as \sfork.
These problems are solved with compiler generated instrumentation, which is used to track memory references and also is used as SFI to provide isolation.
However, CARAT/CARAT CAKE has a number of limitations, including lack of compatibility with unmodified JIT'ed languages due to the use of SFI and worse performance, since CARAT's evaluation shows acceptable overhead only when virtual-to-physical mappings do not change, which is not the case for \fork. 
Memory reference identification is also a problem faced by dynamic ASLR works including RuntimeASLR~\cite{clone_wars}, TASR~\cite{tasr} and Shuffler~\cite{shuffler_aslr} and garbage collectors which must identify pointers to objects they relocate during compaction~\cite{hardwaregc, gcanalysis, compactinggc, garbagefirst}.

\section{Conclusion}
We present \sfork, a design for true single address space \fork using modern architectures.
Previous works sacrifice the lightweightness benefits of SASOSes, isolation, and/or transparency.
In contrast, \sfork preserves a single address space design to maintain lightweightness, uses intra-address-space isolation mechanisms to segregate \processes and the kernel, and works on unmodified applications.
\sfork outperforms a monolithic OS, CheriBSD, as well virtualized SASOS competitor, Nephele, on key \fork/SASOS metrics by an order of magnitude.
Our \sfork prototype is open source and can be found here: \url{https://github.com/flexcap-project/ufork}.
\section{Acknowledgments}
This paper is dedicated to the memory of Gilles Muller.
We thank the anonymous reviewers and our shepherd, Andrew Baumann, for their insightful feedback and invaluable help increasing the paper's quality.
We also thank Binoy Ravindran as well as Alain Tchana and his team for their help in our early efforts to tackle the isssue of supporting \texttt{fork} in a single address space.
Finally, we thank Mikel Luj\'an for his insights and guidance.
This work was partly funded by the UK's EPSRC grant EP/V012134/1 (UniFaaS), EP/V000225/1 (SCorCH), EP/X015610/1 (FlexCap), as well as the grant 10017512 (MoatE) of the UKRI ISCF DSbD Programme.
We acknowledge the support of the Natural Sciences and Engineering Research Council of Canada (NSERC).
Nous remercions le Conseil de recherches en sciences naturelles et en génie du Canada (CRSNG) de son soutien.

\bibliographystyle{ACM-Reference-Format}
\bibliography{references.bib}


\end{document}